\documentclass[]{interact}

\usepackage{epstopdf}
\usepackage[caption=false]{subfig}

\usepackage{tcolorbox} 
\usepackage{multirow} 
\usepackage{multicol} 
\usepackage{threeparttable}
\usepackage{scalefnt}
\usepackage{url}

\usepackage[natbibapa,nodoi]{apacite}
\theoremstyle{plain}

\theoremstyle{definition}

\theoremstyle{remark}

\begin{document}

\articletype{REVIEW ARTICLE}

\title{Example-Based Learning in Software Engineering Education: A Systematic Mapping Study}
\author{
\name{T. P. Bonetti\textsuperscript{a}, \thanks{CONTACT T. P. Bonetti. Email: tiagopiperno@gmail.com}  W. Silva\textsuperscript{b} and T. E. Colanzi\textsuperscript{a}}
\affil{\textsuperscript{a}Universidade Estadual de Maringá (UEM), Maringá, Paraná, Brazil; \textsuperscript{b} Universidade Federal do Pampa (UNIPAMPA), Alegrete, Rio Grande do Sul, Brazil}
}

\maketitle


\begin{abstract}
\textbf{Background and Context}: The discipline of Software Engineering (SE) allows students to understand specific concepts or problems while designing software. Empowering students with the necessary knowledge and skills for the software industry is challenging for universities. One key problem is that traditional methodologies often leave students as passive agents, limiting engagement and learning effectiveness. To address this issue, instructors must promote active learning in the classroom. Among the teaching methodologies, Example-Based Learning (EBL) has shown promise in improving the quality of Software Engineering Education (SEE). \textbf{Objective}: This study aims to investigate and classify the existing empirical evidence about using EBL in SEE. 
\textbf{Method}: We carried out a systematic mapping to collect existing studies and evidence that describe how instructors have been employing EBL to teach SE concepts. 
\textbf{Findings}: By analyzing 30 studies, we identified the benefits and difficulties of using EBL, the SE contents taught by instructors, and the artifacts that support the methodology’s use in the classroom. Besides, we identified the main types of examples used in SEE through EBL. 
\textbf{Implications}: We realized that EBL contributes to student learning, helping in students’ interaction, interpreting and applying concepts, and increasing student motivation and confidence. However, some barriers to adopting EBL in SEE are increasing the effort required by instructors, lack of adequate learning support, and time spent constructing diagrams with errors. Overall, our findings suggest that EBL can improve the effectiveness of SEE, but more research is needed to address the gaps and challenges identified in our study.
\end{abstract}


\section{Introduction}
\label{sec:introducao}

Software Engineering aims to support all aspects of software production, applying the necessary theories, methods, and tools to achieve this goal \citep{sommerville2011software}. As a discipline closely linked to day-to-day software development processes \citep{Marques2014}, students must prepare themselves to effectively apply the concepts they learn during practice \citep{Jianguo2009}. However, optimizing Software Engineering Education (SEE) is a significant challenge for universities because most still adopt traditional teaching methodologies that focus on lectures, leaving the student as a passive agent of learning \citep{cunha2018, Fonseca2017, Ferreira2018}.

Instructors can minimize this problem by adapting or employing new teaching methodologies. In this direction, the literature indicates that a promising and effective approach is Active Learning Methodologies \citep{Lima2021, Fonseca2017, Jianguo2009}. These methodologies use a student-centered teaching strategy to ensure students are actively engaged and involved in learning. Besides, Active Learning Methodologies aim to increase students' interest in the classroom, critical thinking, problem-solving, and decision-making skills, which are fundamental for a software engineer's success \citep{Lima2019}.

Instructors can adopt several Active Learning Methodologies to improve the teaching process in SEE, and Example-Based Learning (EBL) has been particularly effective \citep{van2018example}. This methodology demonstrates how to perform a specific task so students can learn a particular content based on observation \citep{van2018example}. EBL has been used in undergraduate Science, Technology, Engineering, and Mathematics (STEM) courses, including Mathematics \citep{ADAMS2014, renkl2009}, Electrical Circuits \citep{VANGOG2006-trasnfer}, Programming \citep{BEEGE2021}, and Database \citep{Chen2020}. In addition to these areas, some researchers have explored using EBL in teaching Software Engineering and have reported various benefits, such as increasing student motivation \citep{S11_shmallo_constructive_2020}, better learning outcomes \citep{S5_silva_floss_2019}, improved interpretation and application of concepts \citep{S1_silva_is_2017}, and promotion of essential skills for the software industry \citep{S2_tonhao_using_2021}. However, a summary of how EBL is employed by instructors in SEE is still needed.

Therefore, this research aims to investigate and categorize existing empirical evidence on how instructors and the Software Engineering community have been employing EBL to teach Software Engineering concepts, as well as benefits and difficulties. To achieve this, we performed a Systematic Mapping Study (SMS). SMS is a rigorous and structured methodology to identify, select, evaluate, and synthesize relevant literature in a specific research area. It allows researchers to comprehend the current knowledge state on a particular topic, identify gaps and future trends in research \citep{kitchenham2007guidelines}. Thus, we realized the current state of the art and determined how SEE researchers and instructors have used EBL, the Software Engineering contents taught by instructors, the artifacts that support the methodology's use in the classroom, and the benefits and difficulties associated with its use. The main benefits identified are improved interpretation and application of concepts and developing practical skills necessary for the software industry. However, we also observed some barriers to adopting EBL in SEE, e.g., the complexity of understanding the context of the examples and the need for adequate learning support. 

The study is organized as follows: Section \ref{sec:referencialTeorico-ABE} presents the background. Section \ref{sec:materialsAndMethods} describes the systematic mapping's methodology and protocol. Section \ref{sec:results} shows the results. A discussion is presented in Section \ref{sec:discussion}. Finally, conclusions and future work are presented in Section \ref{sec:conclusion}.

\section{Example-Based Learning}
\label{sec:referencialTeorico-ABE}

Active teaching methodologies promote student participation in the construction of their learning, using methods focused on the student's activity, thus contributing to their autonomy \citep{bacich2018}. There are several ways to apply active methodologies, and several studies have pointed to the effectiveness of these methods in teaching Software Engineering-related content \citep{Paez2017, dosSantos2013, Matalonga2017, Ramos2018}.

One active methodology that instructors can adopt is Example-Based Learning (EBL). \cite{van2018example} defines EBL as a methodology that demonstrates how to perform a specific task or how a particular skill can be assimilated for students. When students observe an example of a successfully executed task, their belief in their ability to perform it increases \citep{Huang2017}. For \cite{van2010}, there are two ways to use examples during the student's learning process: Worked Examples and Modeling Examples. To broaden the view of the effects of using examples in the learning process, \citet{Huang2017} adds two more ways in addition to the two already mentioned: Erroneous Worked Examples and Peer Modeling Examples. Figure \ref{fig:tipos-exemplos} shows a classification of examples according to their pedagogical perspective.

[Figure \ref{fig:tipos-exemplos} near here]

\textbf{Worked Examples}, originated from Cognitive Load Theory \citep{Sweller1998}, aims to provide instructions for presenting information to stimulate student behaviors that improve intellectual performance \citep{Sweller1998}. The amount of information that working memory can handle at once is called cognitive load. According to \cite{Sweller1998}, working memory has limited capacity, and instructional methods should avoid overloading it with tasks that do not directly contribute to learning. In this sense, Worked Examples seek to improve learning and subsequent transfer of learning in a different environment, reducing the cognitive load required in working memory \citep{Huang2017}. Besides, worked examples aim to provide students with a didactic procedure for performing a task or solving a problem \citep{Huang2017}. To achieve this, students received a written report with the steps needed to accomplish the goal \citep{van2018example}. \textbf{Erroneous Worked Examples} are considered a variation of Worked Examples where the solution demonstration stage contains errors that students can find and correct \citep{Huang2017, van2010}.

\textbf{Modeling Examples} originate in Social Cognitive Theory, which proposes that we learn from social interactions with others \citep{bandura1977social}. According to \cite{bandura1977social}, individuals acquire new ideas and behaviors by observing others, highlighting the need to observe, model, and imitate behaviors. Modeling Examples are performed through a demonstration, either in person or through videos, by an expert in the skill being demonstrated \citep{van2018example}. \textbf{Peer Modeling Examples} are considered a variation of Modeling Examples, where a student who is not yet considered an expert in the subject has done the modeling and is prone to making errors during the process \citep{Huang2017}.

The EBL in the teaching process explores various fields of knowledge. \cite{ADAMS2014} performed an empirical study using Erroneous Worked Examples to improve math education. The study showed that this kind of example significantly improved students' performance and understanding of deeper problems. \cite{renkl2009} also analyzed the use of worked examples in teaching heuristic domains. They found that examples are helpful in reducing cognitive load and allowing students to focus on the most relevant aspects of the content. The authors stated that using worked examples achieves better knowledge transfer performance with less time and cognitive effort than problem-solving.

In addition to the previously mentioned papers, there are studies evaluating the use of examples in the field of Computing. \cite{BEEGE2021} evaluated using Erroneous Worked Examples in teaching programming. This research analyzed the impact of basic syntactic errors (relating to code structure) and complex semantic errors (relating to code logic or content) in programming. The authors observed that appropriately adjusted syntactic errors reduce mental load and improve student learning. However, they noticed no positive influence on learning regarding semantic errors because students have difficulties detecting and correcting this type of error.

\cite{Chen2020} conducted an empirical study to evaluate the effectiveness of three teaching approaches in database education: problem-solving, worked examples, and erroneous worked examples. The results showed that all approaches significantly improved the level of student learning. However, they noticed that erroneous worked examples were more effective than regularly worked examples because examples with errors require deeper cognitive processing from students. To the authors, this may be related to the fact that examples with errors require deeper cognitive processing from students. They also realized that explaining and correcting examples with errors led students to acquire more advanced debugging and problem-solving skills.

The cited papers demonstrate the effectiveness of using examples in teaching various fields of knowledge, including Computing. The literature also suggests that this methodology can be a good ally in teaching the Software Engineering discipline \citep{S2_tonhao_using_2021, S1_silva_is_2017, S5_silva_floss_2019}. In this sense, teaching Software Engineering using examples is a promising area for further studies in search of more effective ways to use them.

\section{Materials and Methods}
\label{sec:materialsAndMethods}

The objective of this section is to describe the methodology used to carry out the SMS. For this mapping, we followed the guidelines of \cite{kitchenham2007guidelines}. Our SMS was conducted in three phases: planning, execution, and presentation of results. The following subsections present the planning and execution of the study.

\subsection{Objective and Research Questions}

This SMS aims to identify how Example-Based Learning (EBL) is being adopted in teaching Software Engineering in current literature. We defined the following research questions to achieve our objective:

\begin{itemize}

    \item \textbf{RQ1: What types of examples are used in teaching Software Engineering through EBL?} The effectiveness of EBL largely depends on the types of examples provided to students. By identifying the types of examples used, we can better understand which instructional approaches are being adopted and their potential to support learning in Software Engineering.

    \item \textbf{RQ2: What Software Engineering contents are being taught employing EBL?} Understanding the specific Software Engineering topics and contents being taught through EBL is essential to determine how widely this method is being applied across different knowledge areas. This question helps to reveal whether EBL is more effective in certain areas of Software Engineering and how it aligns with industry-relevant skills.

    \item \textbf{RQ3: What artifacts are used in teaching Software Engineering through EBL?} Teaching Software Engineering often involves the use of specific artifacts. Identifying these artifacts is important to assess how they contribute to learning outcomes and to explore the practical tools that support EBL in Software Engineering.

    \item \textbf{RQ4: What type of empirical study has been conducted in papers that assess the employment of EBL in Software Engineering Education (SEE)?} The rigor and type of empirical studies can provide insights into the quality of evidence supporting the use of EBL in SEE. Understanding the types of studies conducted helps to evaluate the strength of the findings in the literature.

    \item \textbf{RQ5: What are the benefits of employing EBL in SEE?} Identifying the benefits of EBL is crucial to understanding its impact on student learning and engagement. This question seeks to highlight how EBL contributes to improved outcomes in SEE.

    \item \textbf{RQ6: What are the difficulties of employing EBL in SEE?} Understanding the difficulties associated with the use of EBL is essential to identify areas that may require adjustments or improvements to ensure a more effective implementation of this educational approach in Software Engineering.
    
\end{itemize}


    
    
    
    
    






    
        
        
        
        
        
    

\subsection{Search Strategy} 
\label{subsec:estrategia}

The first step was the selection of digital libraries, which was guided by their relevance in the field of Computer Science research, their support for Boolean search expressions, and their comprehensive coverage of high-quality publications. We selected four major digital libraries for the search: ACM Digital Library, IEEE Xplore, Scopus, and Springer. ACM and IEEE are recognized as two of the leading resources in the field of Computer Science and Software Engineering, offering a vast collection of peer-reviewed articles and conference proceedings. Scopus, serving as a multidisciplinary metalibrary, indexes publications from renowned publishers such as Elsevier and Springer, providing extensive coverage across multiple disciplines. Springer, in turn, is known for its high-impact journals and conference proceedings in various subfields of Computer Science. The choice of these libraries was based on their prominence in the field, their frequent inclusion in previous systematic reviews \citep{dyba2007applying, petersen2015guidelines, mendes2020when, nakamura2022factors}, and their robust search functionalities, ensuring that we retrieved a diverse and relevant set of studies for this SMS.



The search mechanisms available in most digital libraries are based on textual search expressions. Thus, defining an effective query string is essential for ensuring the quality of the search process and, consequently, the success of this systematic mapping. The query string used for the automatic search in this study was developed based on two main categories derived from the research questions: (1) Example-Based Learning (EBL) and (2) Software Engineering Education (SEE).

Initially, we identified relevant keywords and synonyms for both categories by reviewing related studies and previous research. For Example-Based Learning, the following terms and their synonyms were identified:

\begin{itemize}
    \item \textbf{example-based learning:} example-driven learning, example-driven, learning by examples, learning from errors, learning from examples, positive example, positive examples, worked examples, worked-out examples, anti-patterns, erroneous examples, learning with errors problem, negative example, and negative examples.  
\end{itemize}

For Software Engineering Education, the identified keywords included the following:

\begin{itemize}
    \item \textbf{software engineering education:} software engineering instruction, software engineering projects, software engineering capstone, software development capstone, software engineering course, software development course, software development education, software practice education, teaching software development, and teaching software engineering. 
\end{itemize}

The keywords were iteratively refined by testing them against a control set of papers to ensure comprehensive coverage without retrieving an excessive number of irrelevant papers. Throughout this refinement process, we followed a strategy similar to that proposed by \cite{Zhang2010}, wherein additional search terms were included based on previously selected papers. Moreover, the control papers were continuously tested to ensure that the search strings returned relevant results. After multiple iterations, we reached a satisfactory balance between the breadth of coverage and the number of retrieved papers.

The final query string used in this study is presented below:

\begin{tcolorbox}
\centering
(``example-based learning" \textbf{OR} ``example-driven learning" ``example-driven" \textbf{OR} ``learning by examples" \textbf{OR} ``learning from errors" \textbf{OR} ``learning from examples" \textbf{OR} ``positive example" \textbf{OR} ``positive examples" \textbf{OR} ``worked examples" \textbf{OR} ``worked-out examples" \textbf{OR} ``anti-patterns" \textbf{OR} ``erroneous examples" \textbf{OR} ``learning with errors problem" \textbf{OR} ``negative example" \textbf{OR} ``negative examples") \textbf{AND} (``software engineering education" \textbf{OR} ``software engineering instruction" \textbf{OR} ``software engineering projects" \textbf{OR} ``software engineering capstone" \textbf{OR} ``software development capstone" \textbf{OR} ``software engineering course" \textbf{OR} ``software development course" \textbf{OR} ``software development education" \textbf{OR} ``software practice education" \textbf{OR} ``teaching software development" \textbf{OR} ``teaching software engineering")
\end{tcolorbox}

\subsection{Criteria for Papers Selection}

The papers selected for this systematic mapping study (SMS) were evaluated according to specific criteria that align with the research objectives. Following the guidelines proposed by \cite{Kuhrmann2017}, we established the following inclusion and exclusion criteria to ensure the relevance and quality of the selected articles.

The inclusion criteria (IC) for selecting papers were as follows:

\begin{itemize}
    \item (IC1) Papers that report difficulties and lessons learned of employing Example-Based Learning in SEE;
    \item (IC2) Empirical Papers discussing teaching Software Engineering employing Example-Based Learning;  
    \item (IC3) Papers that employ Example-Based Learning in SEE; 
    \item (IC4) Papers that discuss aspects related to the teaching process using Example-Based Learning in SEE; and,
    \item (IC5) Papers that present technologies that support the teaching of Software Engineering using Example-Based Learning. 
\end{itemize}

Papers that met at least one of the following exclusion criteria (EC) were excluded:

\begin{itemize}
    \item (EC1) Papers that do not employ Example-Based Learning in the context of Software Engineering;
    \item (EC2) Papers that are not written in English;
    \item (EC3) Calls for papers for events or journals;
    \item (EC4) Papers of the following types: books, doctoral theses, master's dissertations, patents, and posters;
    \item (EC5) Papers that do not address the employment of Example-Based Learning in higher education; and,
    \item (EC6) Duplicate Papers.
\end{itemize}





\subsection{Study Selection Process}

The query string presented in Subsection \ref{subsec:estrategia} was used to retrieve candidate papers from the digital libraries in two stages. The first search was conducted in January 2023, and an updated search was performed in January 2024 to ensure that the most recent studies were included in the analysis. The selection process was iterative and incremental, and each paper went through two distinct filtering stages.

Figure \ref{fig:fasesRS} illustrates the process of selecting papers for this SMS. This process began with 814 papers. In the first stage, duplicate papers were located and removed, leaving 745 papers. In the second stage, inclusion (IC) and exclusion (EC) criteria were applied by reading the papers' titles, abstracts, and keywords. At this stage, the selection process was conducted by one researcher and reviewed by two other researchers to ensure consistency and accuracy. At the end of this stage, 55 papers remained. Finally, in the third stage, the inclusion and exclusion criteria were reapplied during the full-text reading of the 55 papers. This stage was also conducted by one researcher and reviewed by the other two. At the end of the process, 30 papers were selected to be included in this SMS.

[Figure \ref{fig:fasesRS} near here]




\subsection{Extraction Process}

We thoroughly read the 30 selected papers to carry out the extraction process. We defined the items for systematic data extraction to obtain and classify the information that would help us answer the research questions. Table \ref{tab:campos_extracao} shows the data extracted from the papers. The first column refers to the aspect of the subject matter in question, and the second column represents the data to be extracted from the work. 

[Table \ref{tab:campos_extracao} near here]

\subsection{Data Analysis}

\begin{figure}
    \centering
    \includegraphics[scale=0.9]{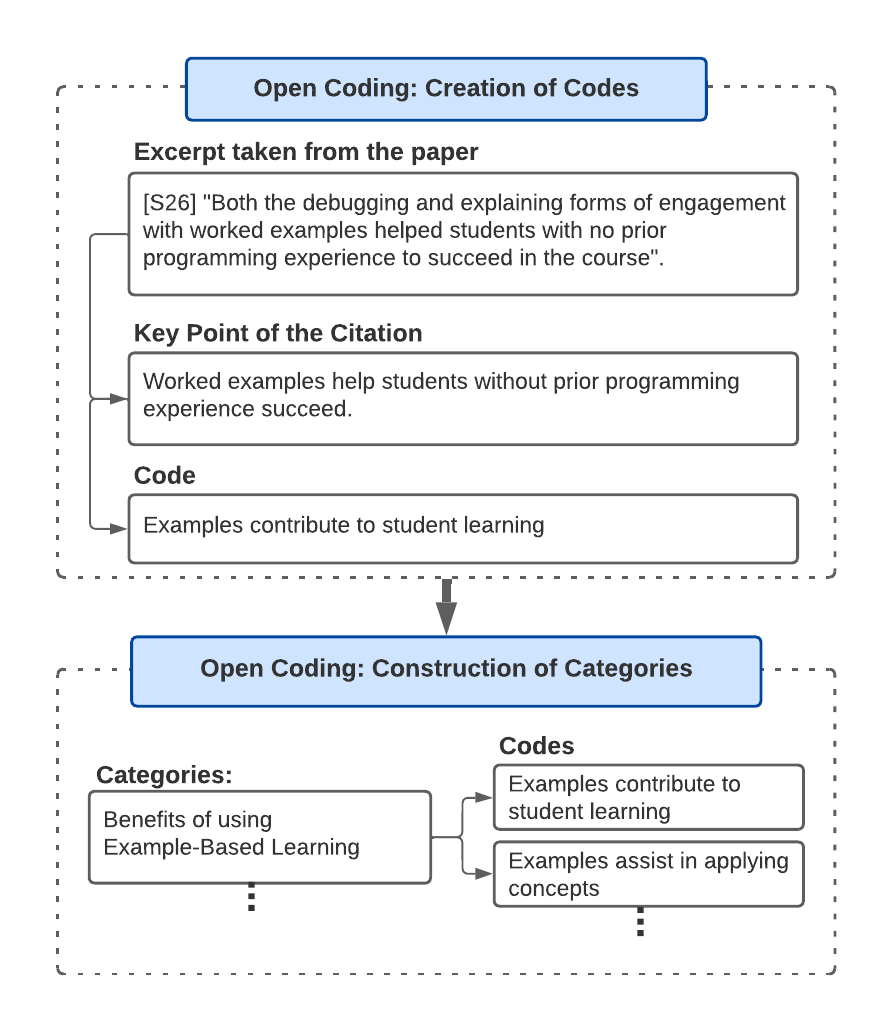}
    \caption{Example of Coding.} 
    \label{fig:codificacao}
\end{figure}

The data analysis followed the open coding procedures described by \cite{strauss_basics_1998}. The goal of this analysis was to organize and synthesize the data extracted from the selected studies to identify the key points regarding the use of EBL in SEE. Initially, we conducted an open coding process, as illustrated in \ref{fig:codificacao}. This process involved breaking down the data into distinct codes and then grouping them into categories to identify common themes across the selected studies. The analysis was carried out by one researcher and subsequently discussed with the rest of the team to mitigate potential biases during the coding process. Further details about the qualitative findings are presented in the following sections.

[Figure \ref{fig:codificacao} near here]

\section{Results}
\label{sec:results}

The following subsections present and discuss the results of this SMS.

\subsection{Overview of Data Extraction}

Figure \ref{fig:estudosPorAno} shows the distribution of the 30 selected studies by year of publication. We can observe that the studies were published between 1998 and 2023. Between 1999 and 2007, there were no publications on the employment of examples to support the teaching of Software Engineering. It is noted that until 2016, the number of published papers remained at one or none per year. From 2017, there were at least two papers per year until 2022, with a notable peak in 2019, when five papers were published. In 2023, there was a further increase, with six studies published throughout the year. This reflects growing interest in the use of Example-Based Learning to support SEE. The total number of papers suggests that, while the topic is gaining traction, there is still room for further exploration and research.


[Figure \ref{fig:estudosPorAno} near here]

Table \ref{tab:locais} displays the publication venues of the studies found in this SMS. The 30 selected studies were published in 22 different venues, demonstrating a wide dispersion of publication venues. Among the venues with the highest number of publications are the conferences: SBES (5) and ICSE (3). Each of the other conferences and journals published one work, which highlights the diversity of interest in EBL in Software Engineering Education across various academic outlets.


[Table \ref{tab:locais} near here]

Table \ref{tab:trabalhosSelecionados} lists the studies selected by the systematic mapping. The first column shows the work identifier that is used throughout the text. The second column presents the title of the work. The third column presents the year of publication of the work. Finally, in the last column we present the authors of the work.

[Table \ref{tab:trabalhosSelecionados} near here]

The following subsections present the answers to each of the research questions of this SMS.

\subsection{RQ1: What types of examples are used in teaching Software Engineering through EBL?}

We classified the papers found in accordance with the information provided and explained in Section \ref{sec:referencialTeorico-ABE}, which covers:  
Worked Examples, Modeling Examples, Erroneous Worked Examples, and Peer Modeling Examples.

Figure \ref{fig:quantidadeTipoExemplo} illustrates the types of examples employed in the analyzed papers in this SMS. We observed that the majority of the papers utilized Worked Examples (18), an instructional method for teaching Software Engineering-related content. Erroneous Worked Examples ranked second with seven papers. Furthermore, five papers explored the use of both Worked Examples and Erroneous Worked Examples. No papers were found that adopted the Modeling Examples approach. These findings indicate that, while Worked Examples are the most common, the combination of Worked and Erroneous Examples has been explored more frequently than initially observed. This outcome could be attributed to the complexity associated with creating and utilizing Erroneous Examples in Software Engineering, as well as the flexibility offered by combining both approaches.

\begin{tcolorbox}
    \centering
    Worked Examples were the most commonly used for teaching Software Engineering, followed by Erroneous Worked Examples. The combination of both types was explored in several studies, possibly due to the flexibility it offers in addressing different aspects of Software Engineering education.
\end{tcolorbox}

[Figure \ref{fig:quantidadeTipoExemplo} near here]


\subsection{RQ2: What Software Engineering contents are being taught employing EBL?}

Table \ref{tab:conteudos} presents the contents taught and and the corresponding frequency of their utilization in the papers examined. The contents were categorized according to the knowledge areas defined in \cite{SWEBOK2014}. The first column shows the knowledge area, the second column indicates the number of papers that addressed that area, the third column shows the content taught, and the fourth column indicates the number of papers that addressed the content.

We noted that most papers focus on software modeling education with UML, which was identified in 26 instances. The following UML models were covered: Use Case Diagram, Class Diagram, Sequence Diagram, Object Diagram, Communication Diagram, and Component Diagram. Specifically, in the context of Software Requirements, the taught models included Use Case Diagram (S10, S12, S13, S14, and S23) and Overview of Requirements (in S17). Additionally, one work addressed the Overview of Requirements (in S17). In the Software Design domain, the taught models were: Class Diagrams (S03, S05, S11, S12, S14, S18, S23, and S27), Sequence Diagrams (S01, S14, and S23), Activity Diagram (S01 and S12), Object Diagrams (S12 and S18), Communication Diagrams (S12), and Component Diagram (S12). Apart from the UML-related topics, other Software Design subjects were addressed in various papers, including Design Patterns (S08, S21, and S22), Refactoring (S02, S04, and S19), Architectural Anomalies (S02 and S04), Software Architecture (S15), and Interface Patterns (S20). Furthermore, nine papers taught Software Construction content, such as Programming (S07, S09, S19, S23, S24, S25, S26, and S29) and Web Application Development using Frameworks (S06). Finally, three papers specifically explored Software Testing and Quality topics, focusing on Verification and Validation (S16 and S28) and Test Case Creation, Execution, and Analysis (S28).

\begin{tcolorbox}
    \centering
    Software modeling with UML was the most frequently taught content, especially through Use Case Diagrams and Class Diagrams. Additionally, Software Engineering contents related to Software Design and Software Construction were also prevalent in the analyzed studies.
\end{tcolorbox}

[Table \ref{tab:conteudos} near here]

\subsection{RQ3: What artifacts are used in teaching Software Engineering through EBL?}

We cataloged the following artifacts in the studies that adopted the Erroneous Examples approach: activity, sequence, and class diagrams with some error (S01, S03, and S27), programming problem sets with and without errors (S07 and S26), a checklist for inspecting use case diagrams (S10), positive and negative guidelines for interfaces (S20), erroneous examples of class diagrams (S18 and S27), and a peer-production-based e-learning system aimed at transferring knowledge encoded in anti-patterns (S08). Additionally, S30 utilized Eye Movement Modeling Examples (EMMEs) to capture the eye movements of experts performing specific tasks and checklists for code review as guiding artifacts for students. Conversely, in the studies that utilized correct examples, the authors reported the following artifacts: worked examples collected from open-source projects (S02, S05, and S19), worked examples developed by instructors (S23, S24, S25, and S29), code snippets written by experts (S09), real examples and scenarios associated with project development (S16), real industry project examples (S17), correct examples of class diagrams (S18 and S27), design patterns (S21), and code review checklists (S30). S29 also employed a scaffolding metacognitive technique, which included guides for planning, monitoring, and reflection during problem-solving.

Furthermore, we identified some tools that provide support for teaching through examples. Firstly, a portal is designed for cataloging Worked Examples (S04), which facilitates the selection and utilization of examples in SEE. Secondly, a web-based learning application (S06) offers a personalized learning path for students to acquire knowledge about web application frameworks using code examples. Thirdly, an exercise system equipped with individual feedback was identified, enhancing students' interactive learning experience (S12). In S28, a repository of test exercises adapted with the TILE method was used, where students had access to code examples and practical exercises related to software testing. Additionally, S22 introduced DEPTHS (Design Patterns Teaching Help System), a learning environment that integrates various teaching tools and systems. This environment encompasses three tools: software modeling, feedback provision, and collaborative annotation. Additionally, it also has an online repository of software patterns. These components are integrated through a flexible ontology-based underlying structure called LOCO (Learning Object Context Ontologies).

\begin{tcolorbox}
    \centering
    The studies utilized a variety of artifacts to support EBL in Software Engineering education, including diagrams (class, activity, and sequence), programming problem sets, checklists, and real-world project examples. These artifacts were primarily used to aid in the teaching of modeling, design, and code review tasks.
\end{tcolorbox}

\subsection{RQ4: What type of empirical study has been conducted in papers that assess the employment of EBL in SEE?}

Concerning the empirical study type, Table \ref{tab:tipoEstudo} shows the data of the studies analyzed in this SMS. The first column represents the types of conducted studies, the second column indicates the number of studies, and the third column displays the study ID. A predominance of experiments is observed, with 11 studies employing this approach. In second place, nine studies conducted case studies, while four studies employed quasi-experiments. Additionally, four studies did not declare the type of study. Among the remaining studies, one adopted action research, and another presented an experience report.

\begin{tcolorbox}
    \centering
    Experiments were the predominant type of empirical study, followed by case studies and quasi-experiments, indicating a preference for controlled study designs in the evaluation of EBL in SEE.
\end{tcolorbox}

[Table \ref{tab:tipoEstudo} near here]


\subsection{RQ5: What are the benefits of employing EBL in SEE?}

Table \ref{tab:benefits} presents the primary benefits of utilizing Example-Based Learning as reported in the analyzed papers. We can observe that \underline{Examples contribute to student learning} (S01, S02, S05, S06, S11, S14, S17, S18, S21, S23, S24, S25, S26, S28, S29, and S30) was the most frequently mentioned benefit in the papers that employed both correct and erroneous examples, including both types within the same work. For instance, in \cite{S25_nurollahian_improving_2023}, it was reported that \textit{``more than 55\% of students who initially wrote with non-expert structure successfully revised to expert structure when prompted''}, highlighting the improvement in learning. Similarly, \cite{S26_garces_engaging_2023} observed that \textit{``both the debugging and explaining forms of engagement with worked examples helped students with no prior programming experience to succeed in the course''}. In one study, it was reported that after students utilized Erroneous Examples (ErrEx) and Problem-Based Learning (PBL), they noticed: \textit{``a good agreement level was obtained relative to both methods (87\% agreement for PBL and 95\% agreement for ErrEx)''}. In another study conducted by \cite{S5_silva_floss_2019}, the use of correct examples improved the students \textit{``understanding of the concepts, their technical preparation for creating class diagrams for a real software project and contributed to better professional performance in the future.''}

[Table \ref{tab:benefits} near here]

Another benefit highlighted in the studies that employed both correct and erroneous examples within the same work is that \underline{Examples help in student interaction} (S01, S05, S14, S17, S19, and S27). In \cite{S27_bonetti_students_2023}, it was mentioned that \textit{``EBL [Example-Based Learning] increased student engagement and interaction during the learning process.''} Regarding this benefit, \cite{S17_daun_project-based_2016} reported that \textit{``after applying the case example-centric approach in the graduate requirements engineering course, we observed improved communication and lively discussions among students.''} Additionally, in the study by \cite{S14_silva_students_2019}, students expressed that \textit{``[the strategy helped] further learning because of the opportunity to work in teams and to hear different opinions.''}

Several benefits have been observed in studies that utilized either correct or erroneous examples. The first benefit identified is that \underline{Examples help interpret concepts}, as mentioned in papers S01, S05, S06, S12, S27, S28, and S30. Regarding this benefit, \cite{S28_doorn_domain_2023} reported that \textit{``these examples help students not only to understand testing concepts but also to identify good test cases, including corner cases.''} The second benefit is that \underline{Examples assist in applying concepts} (S01, S02, S05, S06, S21, and S25). In \cite{S25_nurollahian_improving_2023}, it was observed that \textit{``over 25\% of students who initially wrote non-expert code could properly edit someone else’s non-expert code to expert structure.''} Additionally, \cite{S1_silva_is_2017} state that \textit{``the use of ErrEx promoted a good perception for the students, especially in terms of remembering, interpreting, and applying concepts.''} The third benefit is that \underline{Examples increase student motivation}, as reported in papers S02, S11, S17, and S27. \cite{S11_shmallo_constructive_2020} noted that \textit{``one of the benefits of our developed activities was increased student motivation.''}

The benefit of \underline{Increased confidence due to the use of examples} (S02, S05, S18, S25, S26, and S29) was mentioned in studies that utilized correct, erroneous, or both types of examples. Regarding this aspect, \cite{S26_garces_engaging_2023} reported that \textit{``debugging and explaining helped students with no prior programming experience to gain confidence in their programming skills.''} Similarly, \cite{S29_shin_effects_2023} noted that \textit{``learners were found to perform better in problem-solving tasks when they had self-confidence, which was promoted through the use of metacognitive scaffolding and WOEs.''} On the other hand, the benefit of \underline{Examples helping to identify modeling errors/problems} (S01, S03, S11, S14, and S27) was cited by studies that employed correct, erroneous, or both types of examples. Regarding this benefit, \cite{S3_balaban_pattern-based_2015} reported the following: \textit{``we found out that an increased awareness of modeling problems improves the identification rate of modeling problems in class diagrams.''}

Finally, two more benefits were reported exclusively in studies that utilized correct examples. The first benefit is that \underline{Examples provide contact exposure to real-world environments} (S02, S05, S16, and S19). This benefit was mentioned by a student in an empirical study conducted by \cite{S19_kussmaul_software_2009}: \textit{``I got some good exposure to open source communities and how they work.''} The second benefit is related to \underline{Improving skills for the job market} (S02, S05, and S19). \cite{S2_tonhao_using_2021} reported that \textit{``about improving skills for the job market, [students] agreed that the worked examples can help in this regard.''}

\begin{tcolorbox}
    \centering
    The main benefits of employing EBL in SEE include enhanced student learning, improved interaction, and better application of concepts. Additionally, examples were shown to increase motivation, boost confidence, and prepare students for real-world environments.
\end{tcolorbox}

\subsection{RQ6: What are the difficulties of employing EBL in SEE?}

Table \ref{tab:dificuldades} presents the main difficulties identified in the selected papers. Among them, two difficulties are equally mentioned in three papers each. The first is \underline{creating diagrams with adequate complexity} (S18, S25, S27). In \cite{S25_nurollahian_improving_2023}, it was reported that \textit{``some students required minimal support to revise their code, while others struggled even with progressively detailed hints.''} In \cite{S18_zayan_effects_2014}, the authors stated: \textit{``the class diagram was significant, with lots of concepts, classes, associations, and constraints to digest.''} \cite{S27_bonetti_students_2023} also pointed out the challenges faced by students in creating diagrams with the appropriate level of complexity.

[Table \ref{tab:dificuldades} near here]

The second difficulty cited in three papers is that \underline{examples increase instructors' effort levels} (S17, S20, S26). Regarding this point, \cite{S17_daun_project-based_2016} noted the necessity for \textit{``a more taxing effort on the part of the instructors to supervise and guide student progress in the undergraduate setting.''} Similarly, \cite{S26_garces_engaging_2023} observed that \textit{``teachers found that students in the create-only condition needed additional scaffolding.''} Additionally, \cite{S20_kotze_dont_2008} highlighted that students working with anti-patterns required more instructor assistance compared to those using patterns.

The \underline{lack of adequate learning support} was identified in two papers (S05, S26). \cite{S5_silva_floss_2019} state that \textit{``we have observed that providing adequate learning support for students using FLOSS projects in the classroom is a challenge as students are unlikely to be familiar with all the details of technology tools.''} Similarly, \cite{S26_garces_engaging_2023} noted that \textit{``the worked examples themselves are no guarantee that students will explore these experts’ solutions effectively.''}

Two additional difficulties were highlighted in papers that used both types of examples. The first was \underline{students needing more instructor support}, reported in two papers (S20, S30). In \cite{S30_hauser_experts_2023}, the authors noted that \textit{``there is still a need to help students to apply the strategies presented in the EMMEs.''} \cite{S20_kotze_dont_2008} also emphasized that \textit{``the students in the anti-patterns asked for more assistance, something that did not happen with the patterns group at all.''} The second difficulty related was that \underline{examples do not aid in learning during modeling}, mentioned in two papers (S01, S14). Regarding this issue, \cite{S1_silva_is_2017} reported that students stated that \textit{``ErrEx did not help in learning during modeling.''} Another student in \cite{S14_silva_students_2019} added: \textit{``I already had questions and the negative examples did not help me, because I did not know to distinguish what was right from what was wrong (during the modeling).''}

Additionally, \underline{reduced interaction between students} was cited as a challenge in two papers (S01, S27). \cite{S27_bonetti_students_2023} observed that \textit{``although useful, ABE did not promote increased connection among students.''} Finally, the issue of \underline{identifying errors in diagrams} was mentioned in one paper (S01), as reported by \cite{S1_silva_is_2017}: \textit{``the erroneous examples might have been considered the most difficult due to two factors: (a) the addition of more than one step during the teaching process; and (b) the difficulty in identifying errors in the diagrams.''}

Two more challenges were cited by a single paper each: \underline{understanding the context of the example} (S05) and \underline{the time required to elaborate diagrams with errors} (S14). As mentioned by \cite{S5_silva_floss_2019}, \textit{``using a FLOSS project only for teaching the class diagram of the UML may not have generated the expected results because the students focused more on getting to know GitHub and doing the initial recognition of the FLOSS project that is dedicated to the correct use of syntax and semantics of the notation presented.''} The difficulty in preparing diagrams with errors was detailed by \cite{S14_silva_students_2019}, who noted: \textit{``although the instructor did not like having to prepare the drawings for the problems, s/he added: it’s not a lot of effort if we think that we usually spend an almost whole day to prepare a two-hour class, so it was not that much effort.''}

\begin{tcolorbox}
    \centering
    The primary difficulties of employing EBL in SEE include the increased effort required from instructors and the challenge of creating sufficiently complex examples. Additionally, some students required more support, and identifying errors in diagrams proved to be difficult for learners.
\end{tcolorbox}

\section{Discussion}
\label{sec:discussion}

Enhancing the efficacy of Software Engineering Education to foster the development of proficient software engineers poses a significant challenge, as extensively discussed by numerous scholars. A considerable body of literature examines the effectiveness of Example-Based Learning  within this domain. It thoroughly explores its merits and challenges in the pedagogical realm.

Figure \ref{fig:conteudoExemploEstudo} illustrates the relationship between the type of example used, the content taught, and the study type conducted in the literature. In terms of content taught, there is a prevailing inclination towards utilizing Worked Examples for teaching Software Design. This tendency suggests that instructors perceive this methodology as efficient in imparting such content. As for Worked Examples, a more diversified distribution is observed across various taught topics,  indicating greater flexibility in using this type of example for instructional purposes. Regarding the study  type, three papers have not been assessed through experiments, case studies, or any form of empirical study. This highlights the imperative need to evaluate the proposed methodologies in these papers to ascertain their effectiveness in the instructional process with students. However, it is important to note that these papers constitute a small proportion of the overall evaluated corpus, indicating researchers' conscientiousness in assessing their proposed teaching methods. Such scrutiny is pivotal in providing new researchers with a realistic perspective on what methods have already proven effective and what aspects require refinement when using examples for teaching Software Engineering.

[Figure \ref{fig:conteudoExemploEstudo} near here]

The following sections will delve into the study's findings in-depth, shedding light on the advantages and obstacles encountered when employing EBL in SEE. We will also identify gaps in adopting EBL in SEE and provide recommendations for future research.

\subsection{Example-Based Learning using Worked Examples}

Among the studies examined in this SMS, 14 explored the teaching of Software Engineering using Worked Examples. Out of these, five studies specifically focused on using Worked Examples derived from open-source projects (S2, S4, S5, S17, and S19). Notably, using open-source projects, \cite{S5_silva_floss_2019} (S5) conducted an empirical study centered around class diagram modeling in UML. In this study, researchers assisted instructors in selecting suitable free software projects and creating relevant examples for the classes and instructional sessions.
   These include the potential for enhancing students' capacity to engage with real-world projects and fostering the development of crucial skills highly valued in the job market. Working with examples from open-source projects can contribute to cultivating attributes such as proactivity, effective communication, and adeptness in problem-solving.
The development of these skills was further emphasized in studies (S2 and S19), reinforcing that EBL is valuable in achieving these objectives and effectively contributes to nurturing and honing these essential skills.

To provide instructors with valuable resources and  references for incorporating worked examples from open-source software projects into their teaching,  \cite{S4_tonhao_portal_2020} (S4) developed a portal for cataloging and structuring this type of example. This  portal serves as a centralized platform where instructors of Software Engineering can consult, access, and utilize the cataloged examples, significantly reducing  barriers throughout the process. Additionally, another study conducted by \cite{S2_tonhao_using_2021} (S2) utilized the cataloged examples from the portal during the Software Engineering teaching process. The outcomes demonstrated  that using worked examples not only motivated student engagement in the learning process but also instilled a sense of confidence. It also facilitated exposure to real-world environments and fostered students' autonomy in their learning journey.

Other papers have explored using worked examples without relying on open-source projects. One such study focused on software modeling education. \cite{S12_krusche_interactive_2020} (S12) introduced an interactive learning method that utilized an online software modeling diagram editor. The method comprised five stages: 1) Theory; 2) Example; 3) Classroom exercises; 4) Feedback; and 5) Reflection. 
    In stage 2, worked examples help students relate the learned concepts to a concrete situation.
Furthermore, worked examples are also used in stage 4, where instructors provide exemplary solutions to the exercises proposed in stage 3 to enrich student feedback.

In addition to papers addressing  topics related to software modeling, several studies focused on software construction knowledge. Two of these studies specifically concentrated on programming. In the first one, \cite{S9_wiese_linking_2019} (S9) presented a strategy that exposes students to code snippets containing common beginner and expert patterns to assess their perceptions regarding code readability and style. This approach facilitated the identification of topics in which students are more likely to use or prefer beginner patterns. \cite{Correa2018} (S6) developed a technique called CME (Concerns, Micro-learning, and Examples) to support the learning of frameworks. This technique establishes a learning path that utilizes interests, micro-learning concepts, and example-based learning to teach application development using frameworks. Examples practically demonstrate how to utilize a framework's components for web applications. The presented results indicate that the technique is useful for teaching web application development in less time compared to other reference materials such as documentation and tutorials.

Software Design is another area of knowledge that has been explored using worked examples. \cite{S15_jamal_comparative_2017} (S15) aimed to showcase how to teach client-server architecture using comparative criticism and example-based learning. In this case, students initiate the process by selecting a client-server project question and proposing a solution to the chosen question. Subsequently, the proposed solution is compared with the correct stored solution. A critique is displayed if the proposed solution does not match the stored solution.

In this particular field of knowledge, \cite{S22_hutchison_semantic_2009} (S22) developed a learning environment called DEPTHS (Design Patterns Teaching Help System). This environment integrates various learning tools for teaching software design patterns. The authors propose a methodology that combines Design-Based Learning, Project-Based Learning, and Engagement Theory. Students are presented with a software design problem that they must solve in a workshop-like fashion. Throughout the problem-solving process, they can access an online repository containing examples of software design patterns, facilitated by DEPTHS. Additionally, students have the opportunity to evaluate their projects as well as their peers' projects. This allows them to learn about alternative solutions and identify potential improvements in their solutions.  Finally, all student projects are publicly published and made available, accompanied by rich semantic contextual metadata that facilitates discovery and reuse.

Finally, regarding Worked Examples, a study proposed a methodology for teaching software quality. \cite{S16_joy_kochumarangolil_activity_2018} (S16) suggested a methodology that includes four pedagogical strategies: Flipped Classroom, Project Role-Play for developing project artifacts, teaching through examples, and seminars for students. 
    In this study, \citeauthor{S16_joy_kochumarangolil_activity_2018} used real-life examples and scenarios to improve the understanding of software quality assurance techniques.

\subsection{Example-Based Learning using Erroneous Worked Examples}

Regarding Erroneous Worked Examples, seven studies were identified that employed this approach. Among these studies, five focused on the teaching of software modeling. \cite{S13_bispo_strategies_2019} (S13) conducted a systematic literature review to identify, collect, and analyze strategies for teaching use case modeling. One of the identified strategies was using anti-patterns, examples of problematic solutions to model \citep{Mohamed2010}. Among the obtained results, six studies employed the anti-pattern strategy, five of which came from the same authors. This indicates the need for other researchers to evaluate this type of approach. Additionally, it was found that anti-patterns help reduce the ambiguity and inconsistency in diagrams. 

In another work, \cite{S10_filipe_improving_2021} (S10) proposed a correlation between the difficulties in modeling use case diagrams and strategies to mitigate these difficulties based on quality criteria. The authors also present a mechanism for guiding the creation of checklists to identify defects in use case models. An experiment was conducted using an anti-pattern strategy in conjunction with this mechanism. 
%
    The correlation established by the authors provides support for the notion that the anti-pattern strategy is effective in mitigating ambiguity and inconsistency in use case diagrams.

The experiment conducted by the authors demonstrates that employing an anti-pattern strategy effectively mitigates the difficulties associated with it. The correlation between the strategy and the difficulties, as presented by the authors, assists instructors in identifying the most appropriate strategies for each objective and helps prevent ineffective strategies.

The study by \cite{S1_silva_is_2017} (S1) compared two active methodologies, Problem-Based Learning and Learning from Erroneous Examples (ErrEx). The authors evaluated these methods in the context of teaching UML diagram modeling, specifically observing the correctness and completeness of the diagrams produced by students.
    The study's findings demonstrated that ErrEx contributes to the creation of more correct and complete diagrams. Furthermore, it was evident that ErrEx promotes a strong understanding and application of concepts and facilitates students interaction.

In a related area of software modeling, \cite{S11_shmallo_constructive_2020} (S11) explored a teaching method aimed at novice designers, wherein students learned how to create more precise and accurate class diagrams by identifying errors in their peers' solutions and learning from these mistakes. According to the authors, while not all of the students' comments on their peers' work were entirely correct or precise, they prompted them to re-evaluate their solutions through a more rigorous and comprehensive examination process.
%
   The presented data show that students made fewer errors after learning from their mistakes. This demonstrates the relevance and potential of incorporating this approach (erroneous examples) in teaching.

Another study relevant to class diagrams was presented by \cite{S3_balaban_pattern-based_2015} (S3). In their research, the authors developed  a catalog of anti-patterns and quality issues in class diagrams to enhance students' ability to detect, recognize, and correct errors in this particular type of diagram. Each pattern in the catalog includes an identification structure, evidence of the problem, and potential solutions for correction.
   The authors argue that exposing students to these anti-patterns improves their proficiency in producing accurate diagrams. The study results indicate that this teaching methodology increased students' awareness of modeling issues, significantly improving their ability to identify errors in class diagrams.
The catalog created by the authors can serve as a valuable pedagogical resource for instructors seeking to implement similar strategies. However,one challenge in utilizing erroneous examples in teaching lies in the time required for creating these examples.

In addition to software modeling, studies have focused on other domains that utilize erroneous worked examples. \cite{S7_griffin_designing_2019} (S7) conducted a study on teaching programming with Python language using erroneous worked examples. In this approach, students were presented with practical problems designed to engage them in analyzing well-written example code. These problems encompassed various types: reading, tracing, comparing, explaining, identifying, predicting, and line-by-line tracing. Some problems required students to write code, and some intentionally included bugs. For the experiment, one group of students received a set of problems without any bugs, while another group received problem sets with intentionally introduced bugs. The study's main findings indicated that including intentional errors did not hinder students' learning. The study also contributed to developing 15 design principles for incorporating intentional errors in programming examples.

\cite{S8_leung_ontology_2011} propose a peer-production-based e-learning system for transferring knowledge encoded in software anti-patterns. The system utilizes Webprotege, an ontology editor, and SPARSE, an intelligent system based on ontologies. Combining  these tools creates an extensible system that can be applied to various e-learning scenarios involving anti-patterns. Within this system, students can categorize anti-patterns based on their symptoms or consequences. When selecting an anti-pattern, students are presented with a description and a refactored solution. Students can also access additional information explaining the reasons behind the proposed anti-pattern at this stage. Moreover, students can actively engage with the anti-patterns by contributing their insights and enriching the existing content. This tool provides a valuable resource for gathering and utilizing anti-patterns; however, the authors did not report any experiments on the proposed system.

\subsection{Example-Based Learning using Correct and Erroneous Worked Examples}

In our study, we also came across research investigating Correct and Erroneous Worked Examples. One particular  study by \cite{S18_zayan_effects_2014} (S18) conducted an experiment to assess the effectiveness of Example-Driven Modeling. The researchers examined real loyalty programs in this methodology and developed a class diagram that fulfilled the program requirements. Subsequently, they prepared six partial examples of object diagrams that focused on the key concepts of the model abstractions. These examples encompassed both valid and invalid examples. 
To carry out the activity, the students were provided with an overview of a loyalty program and were required to elaborate an object diagram. Throughout this task, they received the examples created by the researchers. The students positively evaluated the approach, preferring partial examples rather than complete ones, as it allowed them to focus on a small subset of the model at a time. The authors also discovered that this approach resulted in enhancements  in diagram completeness, error reduction, task completeness, and activity efficiency. 
Another noteworthy aspect addressed in the study was the utilization of various example types. According to the authors, 
    one correct example per concept is usually sufficient unless more than one idea is formulating that concept. Erroneous examples are only necessary for constraints that are not easily understandable.

To investigate the impact of active learning strategies on the instruction of UML diagrams, \cite{S14_silva_students_2019} (S14) conducted four case studies employing different methodologies, including  Correct and Erroneous Examples.  The study brings to light significant challenges associated with this methodology. Firstly, the authors observed that students took more time than anticipated to complete the assigned activities. Secondly, the students needed to pay more attention to the provided examples. This underscores the importance of enhancing students' awareness regarding the relevance of these examples and how they can aid in constructing knowledge.

    One challenge mentioned in the literature pertained to instructors' ability to create diagrams with errors. This indicates that the absence of didactic resources related to error examples can diminish instructors' enthusiasm for employing this approach.

Finally, a study by \cite{S20_kotze_dont_2008} aimed to investigate the effectiveness of using erroneous examples instead of correct ones in teaching HCI principles. The participants were divided into two groups: one received positive guidelines and negative guidelines regarding interface patterns. The study findings revealed that students' performance with positive guidelines was significantly better than those taught with negative guidelines.
    Based on these results, it can be inferred that it is not advisable for students to be introduced to the content through erroneous guidelines or examples. Instead, they should be exposed to correct examples that illustrate the appropriate application of the knowledge.

In conclusion, Example-Based Learning has been identified as a promising approach to enhance the effectiveness of teaching Software Engineering and developing competent professionals. However, it is crucial to acknowledge the difficulties and challenges encountered while implementing this approach. The pursuit of effective and tailored solutions for Software Engineering education should persist to ensure the preparation of skilled engineers who can tackle the demands of the job market.

\subsection{Diretrizes para Utilização da ABE no Ensino de Engenharia de Software}

O uso de exemplos trabalhados na educação em Engenharia de Software é uma prática que pode contribuir para uma compreensão mais profunda dos conceitos ensinados e para o desenvolvimento de habilidades práticas pelos estudantes. A seguir, apresentamos diretrizes para a utilização de exemplos no ensino, embasadas nos estudos analisados.


\vspace{0.2cm}
\textbf{1. A seleção dos exemplos é importante para o sucesso do ensino baseado em exemplos.} \citet{S4_tonhao_portal_2020} apontam que o desafio é escolher exemplos que sejam suficientemente ricos para ilustrar problemas reais da Engenharia de Software, mas que não sejam excessivamente complexos a ponto de desmotivar os estudantes. Projetos de Software Livre de porte médio têm se mostrado ideais, pois oferecem a complexidade necessária sem serem inatingíveis. Esses projetos permitem aos alunos compreender os desafios de desenvolvimento, manutenção e colaboração no código, preparando-os melhor para o mercado de trabalho.

Para garantir a adequação dos exemplos, recomenda-se realizar uma avaliação prévia dos projetos disponíveis, considerando fatores como estrutura do código, documentação e relevância dos conceitos abordados. Exemplos que abordam padrões comuns de projeto, bugs reais, refatoramento e funcionalidades amplamente utilizadas em projetos industriais aumentam a relevância e o impacto do ensino. Além disso, a seleção de exemplos deve estar alinhada com os objetivos de aprendizagem específicos de cada disciplina, garantindo que cada exemplo escolhido esteja contribuindo diretamente para o entendimento dos conceitos mais relevantes.


\vspace{0.2cm}
\textbf{2. O uso de anti-padrões é uma estratégia que pode ajudar os alunos a evitar erros recorrentes no desenvolvimento de software.} Conforme \citet{S3_balaban_pattern-based_2015}, anti-padrões são soluções comumente adotadas que, embora populares, não são eficazes e frequentemente levam a problemas de manutenção e qualidade. Apresentar anti-padrões aos estudantes permite que eles reconheçam soluções inadequadas e aprendam as boas práticas para evitá-las.

Essa abordagem pode ser aplicada em atividades teóricas, onde se discutem exemplos de código com anti-padrões, e em práticas, onde os estudantes identificam anti-padrões em códigos reais e propõem melhorias. Essa técnica estimula o pensamento crítico e fortalece o entendimento dos princípios de qualidade em modelagem e implementação de software. Além disso, o estudo de anti-padrões permite que os estudantes desenvolvam uma visão mais crítica sobre as decisões de design e os impactos dessas decisões ao longo do ciclo de vida do software, promovendo uma compreensão mais profunda dos aspectos de manutenção e evolução de sistemas.

\subsection{Erros Intencionais como Ferramenta de Ensino}

A incorporação de erros intencionais é uma técnica eficaz para ensinar modelagem e desenvolvimento em Engenharia de Software \citep{S11_shmallo_constructive_2020}. Consiste em apresentar aos estudantes diagramas ou códigos contendo erros propositalmente incluídos, incentivando-os a identificar e corrigir os erros, o que contribui para uma aprendizagem ativa.

Essa abordagem também pode promover a reflexão crítica sobre o próprio processo de aprendizado, incentivando os estudantes a refletirem sobre seus processos de pensamento e identificação de problemas. Utilizar erros comuns, como relações incorretas em diagramas de classes ou falhas em especificações de requisitos, demonstra como pequenos detalhes podem impactar significativamente um projeto. Essa prática pode ser usada também em atividades colaborativas, onde os estudantes discutem as melhores soluções para corrigir os problemas apresentados. Essa abordagem não apenas pode melhorar a capacidade técnica dos alunos de identificar e corrigir erros, mas também pode desenvolver sua confiança para lidar com problemas inesperados, uma habilidade essencial no ambiente profissional.

\subsection{Uso Construtivo de Bugs}

A incorporação de bugs intencionais em códigos apresentados aos estudantes é uma prática sugerida por \citet{S7_griffin_designing_2019} para desenvolver habilidades de depuração. Resolver problemas contendo erros planejados permite que os alunos explorem cenários reais de depuração, compreendendo os tipos de erros mais comuns e desenvolvendo estratégias para corrigi-los.

Essa atividade também estimula a construção de conhecimento negativo, ou seja, a compreensão do que não deve ser feito e das consequências de soluções inadequadas. Aprender a evitar erros comuns é tão valioso quanto implementar soluções corretas, principalmente quando os estudantes são desafiados a depurar códigos que simulam situações reais. Ao promover atividades de depuração, o professor deve garantir que os estudantes entendam o raciocínio por trás dos erros e as melhores práticas para evitá-los no futuro, aumentando assim a resiliência dos alunos frente a problemas complexos.

\subsection{Exemplos Concretos para Melhor Compreensão de Modelos}

\citet{S18_zayan_effects_2014} mostram que o uso de exemplos concretos durante a modelagem facilita a compreensão dos estudantes sobre conceitos abstratos, melhorando a completude dos diagramas e reduzindo erros. Esses exemplos fornecem uma representação prática dos conceitos, permitindo que os alunos vejam como as ideias teóricas se conectam com implementações práticas e visualizem as interações entre diferentes componentes do sistema.

O uso de exemplos explícitos ao introduzir novos conceitos de modelagem pode promover discussões mais profundas e incentivar os alunos a explorar os conceitos em maior profundidade. Isso pode contribuir para uma melhor compreensão e desenvolver habilidades analíticas críticas, ajudando a reduzir a carga cognitiva inicial e tornando o aprendizado mais acessível. Adicionalmente, exemplos concretos podem ajudar a estabelecer uma ponte entre teoria e prática, especialmente quando se trata de conceitos abstratos e de difícil visualização, o que pode melhorar significativamente a retenção do conhecimento e sua aplicabilidade.

\section{Threats to Validity}
\label{sec:ThreatsToValidity}

Although the protocol for this SMS was carefully defined following \cite{kitchenham2007guidelines} and systematically applied, this research still faces several well-known limitations and threats to its validity. However, several strategies were employed to mitigate the impact of these factors, focusing on the construction of the search string, study selection, and data extraction.

As highlighted by \cite{Ampatzoglou2019}, a common threat in systematic mapping studies is related to the search string construction. To minimize this risk, we carefully developed a comprehensive search string aimed at including all potentially relevant publications. This process involved several iterations, ensuring the search terms covered a wide range of synonyms and variations for EBL and SEE.

In terms of study selection, which is another well-known threat to validity, we mitigated biases by clearly defining rigorous inclusion and exclusion criteria. These criteria were established at the beginning of the study and documented carefully to ensure consistency. Furthermore, the selection process involved two independent researchers who reviewed the studies. Weekly discussions were held to reach a consensus, particularly for studies that did not clearly fit the criteria, reducing the risk of inclusion/exclusion bias. Another prevalent threat to validity is related to data extraction, where biases can occur in interpreting and extracting relevant information from the studies. To address this, we predefined possible answers to each question in the protocol, ensuring a structured extraction process. Data extraction was primarily conducted by the first author, and when information was not explicitly stated in the articles, reasonable inferences were made based on the context. All extracted data were reviewed by the co-authors to ensure accuracy and mitigate potential bias in interpretation.

Finally, the selection of digital libraries poses a potential threat to the completeness of the study. To mitigate this, we chose widely recognized and frequently used digital libraries in the field of computer science, such as IEEE Xplore, ACM Digital Library, and Scopus. This helped ensure that we captured a broad range of relevant studies, minimizing the risk of missing key publications. By carefully addressing these threats, we aimed to enhance the robustness and reliability of the findings from this systematic mapping study, though it is acknowledged that certain limitations remain inherent in the process.
\section{Conclusion}
\label{sec:conclusion}

Software engineers play a fundamental role in developing high-quality software. To ensure these professionals are well-equipped to tackle the profession's challenges, educational institutions must employ teaching methods that facilitate the acquisition of knowledge encompassing the entire software development process. In this sense, this work aimed to explore the application of Example-Based Learning in  Software Engineering Education.

Consequently, this SMS aimed to identify publications focusing on implementing Example-Based Learning in SEE. Based on the articles retrieved, it became evident that this methodology could potentially enhance learning outcomes in the field. The benefits of this method include facilitating the interpretation and application of concepts and fostering the development of essential skills sought after in the job market. However, the studies also acknowledged certain challenges, such as the inherent complexity of comprehending the context of the examples and the requirement for adequate learning support.

Despite advances in research on EBL in SEE, gaps and challenges still need to be addressed. For example, one area that requires further investigation is adapting EBL to different educational and cultural contexts. Exploring and understanding how EBL can be integrated with other teaching methodologies is essential to maximize its effectiveness. Developing educational tools and resources specifically designed to support instructors in implementing example-based learning is another crucial aspect that deserves attention. These tools can provide guidance, facilitate the creation and selection of examples, and offer feedback mechanisms to enhance the learning process. Therefore, future work should focus on filling these gaps and addressing the challenges to make EBL even more effective in SEE.

\section{Acknowledgements}

The authors thank the Coordenação de Aperfeiçoamento de Pessoal de Nível Superior - Brasil (CAPES) - Código Financeiro 001. Williamson Silva thanks FAPERGS (Projeto ARD/ARC -- process 22/2551-0000606-0), and the Universidade Federal do Pampa (UNIPAMPA - Alegrete). Thelma Elita Colanzi Lopes thanks CNPq (Projeto Universal código 408812/2021-4, and CNPq\_80364faece).

\bibliographystyle{apacite}
\bibliography{interactapasample}

\clearpage

\label{sec:tables}

\begin{table}
\caption{Extraction fields}
\small 
\centering

	\label{tab:campos_extracao}
\begin{tabular}{|p{5cm}|p{8cm}|}

\hline
\textbf{Aspect} & \textbf{Data to be extracted} \\ \hline
General information & Work title \\
                   & Author(s) \\
                   & Publication type \\
                   & Publication year \\
                   & Publication venue \\ \hline
Employment of EBL & SRQ1: Type of example used. \\
                   & SRQ2: Benefit(s) identified in the study. \\
                   & SRQ3: Difficulty(ies) identified in the study. \\ \hline
About the taught content & SRQ4: Knowledge area. \\
                   & SRQ5: Taught content(s). \\
                   & SRQ6: Artifact(s) used. \\ \hline
About the study     & SRQ7: Type of study conducted. \\ \hline

\end{tabular}
\end{table}

\clearpage

\begin{table}
\caption{Publication Venues of the Works}
\small 
\centering
	
\label{tab:locais}
	
\begin{tabular}{| p{9.9cm} | p{1.6cm} | c |}
\hline

\textbf{Publication Venue} & \textbf{Type} & {\textbf{Quantity}} \\ \hline
    Brazilian Symposium on Software Engineering (SBES) & Conference   & 5 \\ \hline
    
    International Conference on Software (ICSE) & Conference   & 3 \\ \hline

    Conference on Software Engineering Education and Training (CSEE\&T) & Conference   & 2 \\ \hline
    
    Annals of Software Engineering  & Journal & 1 \\ \hline
    
    ASEE Annual Conference and Exposition, Conference Proceedings & Conference   & 1 \\ \hline

    Computers and Education & Journal & 1 \\ \hline

    Conference on e-Learning, e-Management and e-Services (IC3e) & Conference & 1 \\ \hline
    
    Education and Information Technologies & Journal & 1 \\ \hline

    European Conference on Software Engineering Education (ECSEE) & Conference & 1 \\ \hline
    
    IEEE Access & Journal & 1 \\ \hline    

    IEEE/ACM International Conference on Software Engineering: Software Engineering Education and Training (ICSE-SEET) & Conference   & 1 \\ \hline
    
    International Conference on Enterprise Information Systems (ICEIS) & Conference   & 1 \\ \hline

    International Conference on Web Engineering (ICWE) & Conference   & 1          \\ \hline

    International Conference on Web-Based Learning (ICWL) & Conference & 1       \\ \hline

    International Workshop on Software Engineering Education for the Next Generation (SEENG) & Conference & 1 \\ \hline    

    Journal of Educational Computing Research & Journal & 1 \\ \hline

    Journal of Information Systems Education & Journal & 1                    \\ \hline
    
    Journal of Information Technology Education: Innovations in Practice & Journal & 1  \\ \hline
    
    Research Challenges in Information Science: Information Science and the Connected World (RCIS) & Conference & 1 \\ \hline
    
    Software and Systems Modeling (SoSyM) & Journal & 1 \\ \hline
    
    The Semantic Web (ISWC) & Conference & 1 \\ \hline

    Transactions on Learning Technologies (TLT) & Journal & 1 \\ \hline 
    
    UK \& Ireland Computing Education Research (UKICER) & Conference  & 1 \\ \hline
\end{tabular}
\end{table}

\clearpage

\begin{table}
\small
\centering
	\caption{Selected works}
	\label{tab:trabalhosSelecionados}

\begin{tabular}{|l|p{8.5cm}|l|p{3.6cm}|}
\hline
\textbf{ID} & \textbf{Title of Selected Work} & \textbf{Year} & \textbf{Authors} \\ \hline

S01 & Is It Better to Learn from Problems or Erroneous Examples? & 2017 & \cite{S1_silva_is_2017} \\ \hline

S02 & Using Real Worked Examples to Aid Software Engineering Teaching & 2021 & \cite{S2_tonhao_using_2021} \\ \hline

S03 & A pattern-based approach for improving model quality & 2015 & \cite{S3_balaban_pattern-based_2015} \\ \hline

S04 & A portal for cataloging worked examples extracted from open source software & 2020 & \cite{S4_tonhao_portal_2020} \\ \hline

S05 & FLOSS in Software Engineering Education: Supporting the Instructor in the Quest for Providing Real Experience for Students & 2019 & \cite{S5_silva_floss_2019} \\ \hline

S06 & CME -- A Web Application Framework Learning Technique Based on Concerns, Micro-Learning and Examples & 2018 & \cite{S6_mikkonen_cme_2018} \\ \hline

S07 & Designing Intentional Bugs for Learning & 2019 & \cite{S7_griffin_designing_2019} \\ \hline

S08 & An Ontology Based E-Learning System Using Antipatterns & 2011 & \cite{S8_leung_ontology_2011} \\ \hline

S09 & Linking Code Readability, Structure, and Comprehension Among Novices: It’s Complicated & 2019 & \cite{S9_wiese_linking_2019} \\ \hline

S10 & Improving Quality of Use-Case Models by Correlating Defects, Difficulties, and Modeling Strategies & 2021 & \cite{S10_filipe_improving_2021} \\ \hline

S11 & Constructive Use of Errors in Teaching the UML Class Diagram in an IS Engineering Course & 2020 & \cite{S11_shmallo_constructive_2020} \\ \hline

S12 & An interactive learning method to engage students in modeling & 2020 & \cite{S12_krusche_interactive_2020} \\ \hline

S13 & Strategies for Use Case Modeling: A Systematic Literature Review & 2019 & \cite{S13_bispo_strategies_2019} \\ \hline

S14 & Students’ and Instructors’ Perceptions of Five Different Active Learning Strategies Used to Teach Software Modeling & 2019 & \cite{S14_silva_students_2019} \\ \hline

S15 & Comparative critiquing and example-based approach for learning client-server design & 2017 & \cite{S15_jamal_comparative_2017} \\ \hline

S16 & Activity Oriented Teaching Strategy for Software Engineering Course: An Experience Report & 2018 & \cite{S16_joy_kochumarangolil_activity_2018} \\ \hline

S17 & Project-Based Learning with Examples from Industry in University Courses: An Experience Report from an Undergraduate Requirements Engineering Course & 2016 & \cite{S17_daun_project-based_2016} \\ \hline

S18 & Effects of using examples on structural model comprehension: a controlled experiment & 2014 & \cite{S18_zayan_effects_2014} \\ \hline

S19 & Software projects using free and open-source software: Opportunities, challenges, and lessons learned & 2019 & \cite{S19_kussmaul_software_2009} \\ \hline

S20 & Don’t do this – Pitfalls in using anti-patterns in teaching human–computer interaction principles & 2008 & \cite{S20_kotze_dont_2008} \\ \hline

S21 & Utilizing patterns and pattern languages in education & 1998 & \cite{S21_kendall_utilizing_1998} \\ \hline

S22 & Semantic Web Technologies for the Integration of Learning Tools and Context-Aware Educational Services & 2009 & \cite{S22_hutchison_semantic_2009} \\ \hline

S23 & Aligning the learning Experience in a Project-Based Course: lessons learned from the Redesign of a Programming Lab & 2022 & \cite{S23_Aligning_the_learning_2022}   \\ \hline

S24 & Collaborative Programming for Work-Relevant Learning: Comparing Programming Practice With Example-Based Reflection for Student Learning and Transfer Task Performance & 2022 & \cite{S24_Collaborative_Programming_2022} \\ \hline

S25 & Improving Assessment of Programming Pattern Knowledge through Code Editing and Revision & 2023 & \cite{S25_nurollahian_improving_2023} \\ \hline

S26 & Engaging students in active exploration of programming worked examples & 2023 & \cite{S26_garces_engaging_2023} \\ \hline

S27 & Students’ Perception of Example-Based Learning in Software Modeling Education & 2023 & \cite{S27_bonetti_students_2023} \\ \hline

S28 & Domain TILEs: Test Informed Learning with Examples from the Testing Domain & 2023 & \cite{S28_doorn_domain_2023} \\ \hline

S29 & The Effects of Worked-Out Example and Metacognitive Scaffolding on Problem-Solving Programming & 2023 & \cite{S29_shin_effects_2023} \\ \hline

S30 & The Expert’s View: Eye Movement Modeling Examples in Software Engineering Education & 2023 & \cite{S30_hauser_experts_2023} \\ \hline

\end{tabular}
\end{table}


	

        

\begin{table}
\small
\centering
\caption{Taught contents}
	\label{tab:conteudos}
	
\begin{tabular}{|l|c|p{4.5cm}|c|}
\hline
\textbf{Area of Knowledge} & \textbf{Quantity} & \textbf{Software Engineering Content} & \textbf{Quantity} \\ \hline

\multirow{3}{*}{Software Requirements} & \multirow{3}{*}{6}   
        & Use Case Diagram & 5 \\ \cline{3-4}      
        & & Overview of Requirements & 1 \\ \hline 
        
\multirow{12}{*}{Software Design} & \multirow{12}{*}{28} 
        & Class Diagram & 8 \\ \cline{3-4} 
        & & Design Patterns & 3 \\ \cline{3-4}   
        & & Refactoring & 3 \\ \cline{3-4}   
        & & Sequence Diagram & 3 \\ \cline{3-4}
        & & Activity Diagram & 2 \\ \cline{3-4}
        & & Object Diagram & 2 \\ \cline{3-4}   
        & & Architectural Anomalies & 2 \\ \cline{3-4}   
        & & Communication Diagram & 1 \\ \cline{3-4}   
        & & Component Diagram & 1 \\ \cline{3-4}   
        & & Interface Patterns & 1 \\ \cline{3-4}   
        & & Software Architecture & 1 \\ \cline{3-4}   
        & & UML Diagrams (unspecified) & 1 \\ \hline 
        
\multirow{2}{*}{Software Construction} & \multirow{2}{*}{9}   
        & Programming & 8 \\ \cline{3-4}   
        & & Web Application Frameworks & 1 \\ \hline 
        
\multirow{2}{*}{Software Testing} & \multirow{2}{*}{3}   
        & Verification and Validation & 2 \\ \cline{3-4}   
        & & Test Case Creation, Test Execution, Result Analysis, etc. & 1 \\ \hline

\end{tabular}
\end{table}

\clearpage

	
    

\begin{table}
\small 
\centering
\caption{Type of study}
	\label{tab:tipoEstudo}
	
\begin{tabular}{|l|c|l|}
\hline
\textbf{Type of study} & \textbf{Quantity} & \textbf{Works} \\ \hline
    Experiment & 11 & S01, S03, S10, S18, S20, S24, S25, S26, S28, S29, S30 \\ \hline
    Case Study & 9 & S02, S05, S07, S09, S11, S14, S17, S19, S22 \\ \hline
    Quasi-experiment & 4 & S04, S06, S12, S27 \\ \hline
    Undeclared & 4 & S08, S13, S15, S21 \\ \hline
    Action Research & 1 & S16 \\ \hline
    Experience Report & 1 & S23 \\ \hline
\end{tabular}
\end{table}

\clearpage

\begin{table}
\small 
\begin{threeparttable}[b]
\centering
\caption{Benefits of using Example-Based Learning}
\label{tab:benefits}

\begin{tabular}{|p{7.3cm}|p{6.2cm}|}
    \hline
    \textbf{Benefits}                                  & \textbf{References} \\ \hline
    
    Examples contribute to student learning           & S01\tnote{2}, S02\tnote{1}, S05\tnote{1}, S06\tnote{1}, S11\tnote{2}, S14\tnote{3}, S17\tnote{1}, S18\tnote{3}, S21\tnote{1}, S23\tnote{1}, S24\tnote{1}, S25\tnote{1}, S26\tnote{3}, S27\tnote{3}, S28\tnote{1}, S29\tnote{1}, S30\tnote{1} \\ \hline
    
    Examples assist in applying concepts               & S01\tnote{2}, S02\tnote{1}, S05\tnote{1}, S06\tnote{1}, S21\tnote{1}, S25\tnote{1} \\ \hline
    
    Examples help in student interaction               & S01\tnote{2}, S05\tnote{1}, S14\tnote{3}, S17\tnote{1}, S19\tnote{1} \\ \hline
    
    Examples help in identifying modeling errors/problems  & S01\tnote{2}, S03\tnote{2}, S11\tnote{2}, S14\tnote{3}, S27\tnote{3} \\ \hline
    
    Examples help interpret concepts                   & S01\tnote{2}, S05\tnote{1}, S06\tnote{1}, S12\tnote{1}, S27\tnote{3}, S28\tnote{1}, S30\tnote{1} \\ \hline
    
    Examples provide contact exposure to real-world environments & S02\tnote{1}, S05\tnote{1}, S16\tnote{1}, S19\tnote{1} \\ \hline
    
    Examples increase student motivation               & S02\tnote{1}, S11\tnote{2}, S17\tnote{1}, S27\tnote{3} \\ \hline
    
    Increased confidence due to the use of examples    & S02\tnote{1}, S05\tnote{1}, S18\tnote{3}, S25\tnote{1}, S26\tnote{3}, S27\tnote{3}, S29\tnote{1} \\ \hline
    
    Improving skills for the job market                & S02\tnote{1}, S05\tnote{1}, S19\tnote{1} \\ \hline
    
\end{tabular}

\begin{tablenotes}
\scriptsize
   \item [1] Worked Examples.
   \item [2] Erroneous Worked Examples.
   \item [3] Both.
\end{tablenotes}
\end{threeparttable}
\end{table}

\clearpage

    
        
    

\begin{table}
\small 
    \centering
    \begin{threeparttable}[b]
    \caption{Difficulties of using Example-Based Learning}
    \label{tab:dificuldades}
    
    \begin{tabular}{|l|p{4.3cm}|}
    \hline
        \textbf{Difficulties}                                & \textbf{Studies} \\ \hline
        
        Creating diagrams with adequate complexity           & S18\tnote{3}, S25\tnote{1}, S27\tnote{3} \\ \hline
        Examples increase instructors' effort levels         & S17\tnote{1}, S20\tnote{3}, S26\tnote{3} \\ \hline
        Lack of adequate learning support                    & S05\tnote{1}, S26\tnote{3} \\ \hline
        Students needing more instructor support             & S20\tnote{3}, S30\tnote{1} \\ \hline
        Examples do not aid in learning during modeling      & S01\tnote{2}, S14\tnote{3} \\ \hline
        Reduced interaction between students                 & S01\tnote{2}, S27\tnote{3} \\ \hline
        Understanding the context of the example             & S05\tnote{1} \\ \hline
        Time required to elaborate diagrams with errors      & S14\tnote{3} \\ \hline
        Identifying errors in the diagrams                   & S01\tnote{2} \\ \hline
        
    \end{tabular}
    
    \begin{tablenotes}
    \scriptsize
       \item [1] Worked Examples.
       \item [2] Erroneous Worked Examples
       \item [3] Both. 
    \end{tablenotes} 
    \end{threeparttable} 
\end{table}

\clearpage
\label{sec:figures}


\begin{figure}
    \centering    
    \includegraphics[scale=0.63]{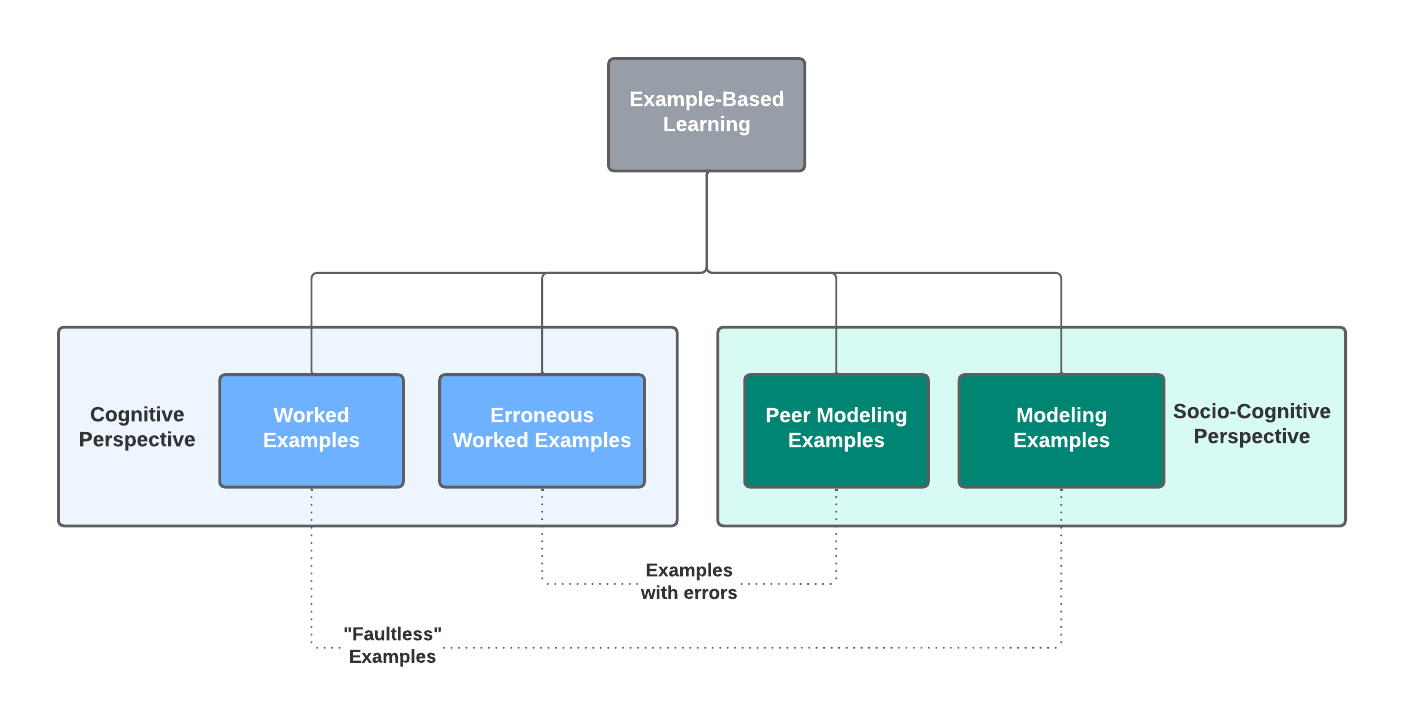}\    
    \caption{Classification of examples in Example-Based Learning context.}    
    \footnotesize Source: Adapted from \cite{Huang2017}
    \label{fig:tipos-exemplos}    
\end{figure}

\begin{figure}
    \centering
    \includegraphics[scale=0.7]{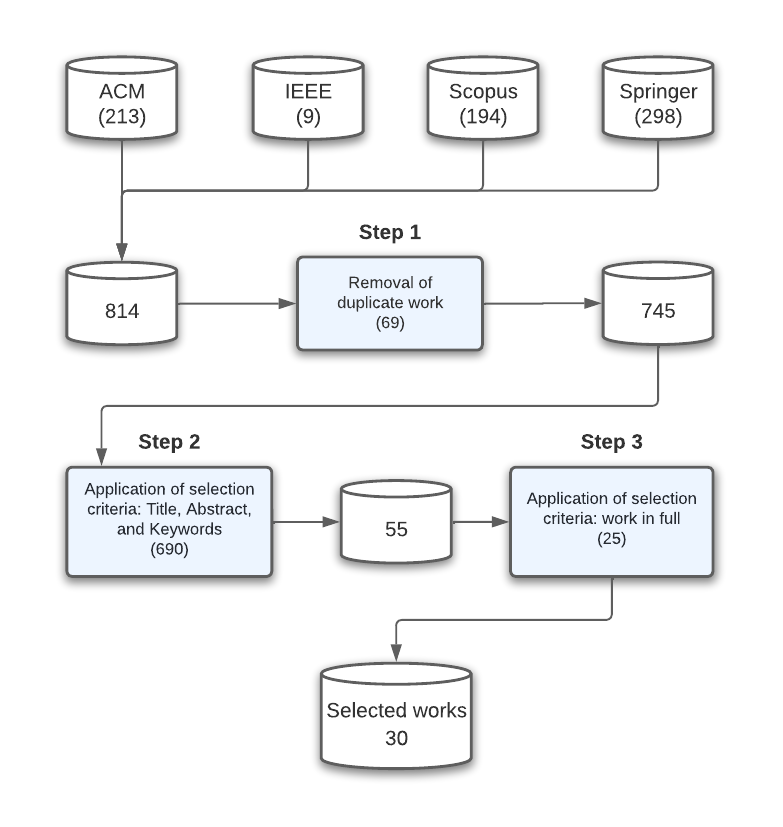}
    \caption{Selection process of the papers.} 
    \label{fig:fasesRS}
\end{figure}

\begin{figure}
\centering
    \includegraphics[scale=0.55]{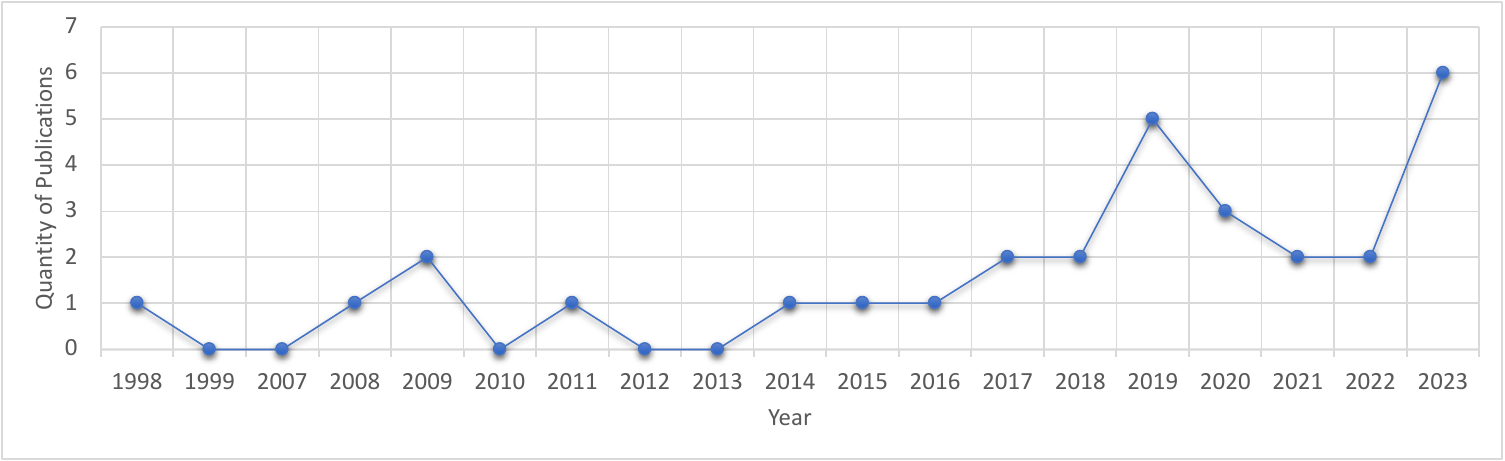}\\
    \caption{Distribution of studies per year}
    \label{fig:estudosPorAno}
\end{figure}

\begin{figure}
\centering
    \includegraphics[scale=0.9]{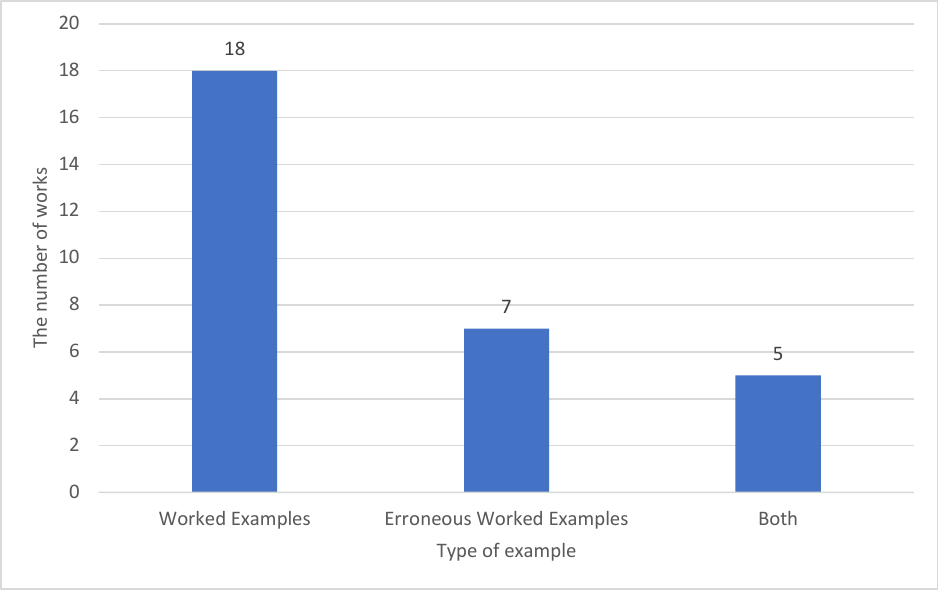}\\
    \caption{Type of Example Used}
    \label{fig:quantidadeTipoExemplo}
\end{figure}

\begin{figure}
    \centering
    \includegraphics[scale=0.50]{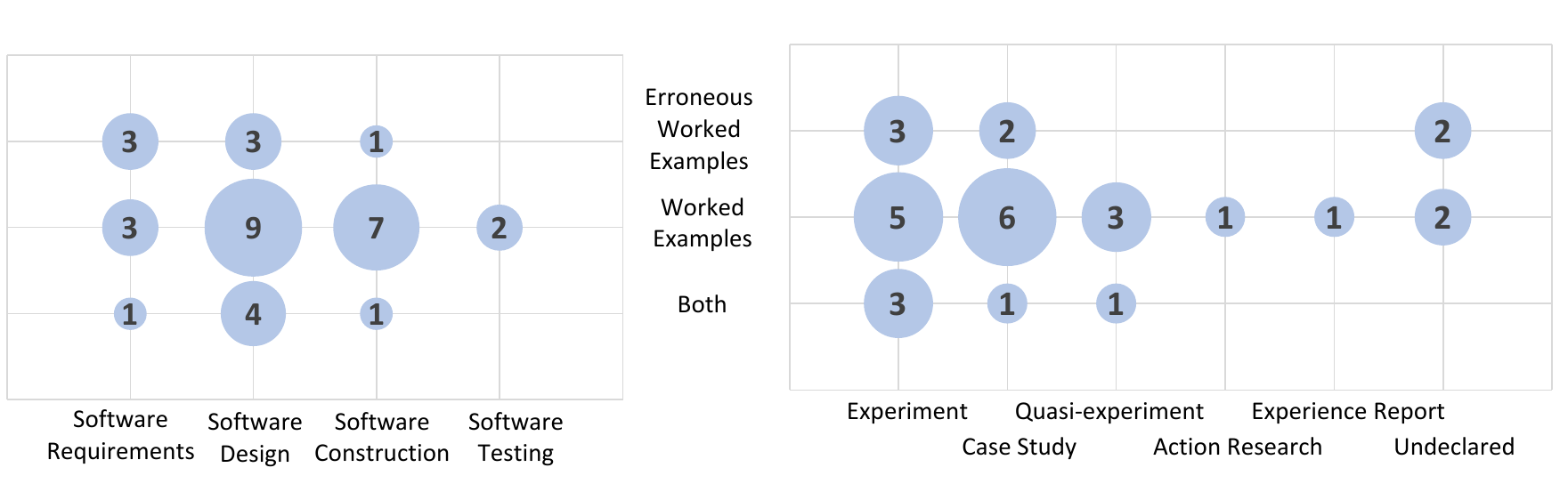}\\
    \caption{Correlation between the content taught and the study conducted by the type of example used}
    \label{fig:conteudoExemploEstudo}
\end{figure}

\label{sec:figuresCaptions}

\clearpage

\textbf{Figure captions}

\begin{itemize}
    \item Figure \ref{fig:tipos-exemplos}: Classification of examples in Example-Based Learning context.
    \item Figure \ref{fig:fasesRS}: Selection process of the papers.
    \item Figure \ref{fig:estudosPorAno}: Distribution of studies per year.
    \item Figure \ref{fig:quantidadeTipoExemplo}: Type of Example Used.
    \item Figure \ref{fig:conteudoExemploEstudo}: Correlation between the content taught and the study conducted by the type of example used.
\end{itemize}


\end{document}